\documentclass[conference]{IEEEtran}
\ifCLASSINFOpdf
\else
  \usepackage[dvips]{graphicx}
  \graphicspath{{./figures/}}
  \DeclareGraphicsExtensions{.eps}
\fi
\begin{document}
%
\title{Bottom-up Broadband Initiatives in the Commons for Europe Project}



%
\author{\IEEEauthorblockN{Jaume Barcelo\IEEEauthorrefmark{1},
Boris Bellalta\IEEEauthorrefmark{1},
Roger Baig\IEEEauthorrefmark{2}, 
Ramon Roca\IEEEauthorrefmark{2},\\
Albert Domingo\IEEEauthorrefmark{1},
Luis Sanabria\IEEEauthorrefmark{1},
Cristina Cano\IEEEauthorrefmark{1} and
Miquel Oliver\IEEEauthorrefmark{1}}
\IEEEauthorblockA{\IEEEauthorrefmark{1} Universitat Pompeu Fabra, email: name.surname@upf.edu}
\IEEEauthorblockA{\IEEEauthorrefmark{2} Gufi.net, email: name.surname@guifi.net}
}


\maketitle

\begin{abstract}
This paper offers an overview of the Commons for Europe (C4EU) project and the role of Bottom-up Broadband (BuB) in developing the information society.
BuB is characterized by the fact that the beneficiaries of the networks actively participate in the planning, deployment and maintenance tasks.
For the beneficiaries, this represent a paradigm shift from a consumer-only position to an active-participant position.

We summarize a representative set of the BuB pilot proposals that have been considered in the context of the C4EU project.
A selection of these proposals will be executed and carefully documented to define good practices in BuB deployments.
The documentation will include project templates, work plans, case studies, replicable success models and regulatory guidelines.

The overall goal of the project is to assess the validity of the BuB model to effectively and efficiently complement exiting traditional deployments in satisfying the networking and technological needs of the European citizens and organizations.

\end{abstract}


%
\IEEEpeerreviewmaketitle

\section{Introduction}
The role of the broadband information networks is gaining importance as the information society develops.
Broadband access means access to a plethora of multimedia educational resources, collaboration opportunities and productivity tools that empower our society.
Ideally, broadband services should be accessible to all the citizens independently of their age, education, economical status or geographical distribution.

With this goal, several community networks have appeared to promote the access to broadband in highly collaborative scenarios.
These community networks are typically led by technology enthusiasts willing to contribute to their community.
To succeed and grow, it is necessary to involve other people that might not be, initially, technologically savvy.
Nevertheless, the high degree of participation and interaction between the participants create the right environment for knowledge transfer and many of the users involved in these community networks quickly acquire advanced networking skills.

It is possible to establish some parallelisms between the community networks and the open source community.
Each community network has its own \emph{developers}, but all the participants in different community networks share a set of principles, good practices and high-level objectives.

Other actors active in the promotion of broadband access are some public institutions, typically municipalities, that have been compelled to offer basic networking services to their citizens and visitors in the form of hotspots.
Even though these services are very limited, they are greatly appreciated by the users.

There is a desire to extend and complement current network deployment efforts, find easily replicable success models and establish collaborations among highly heterogeneous groups that share similar goals.
In this context, Bottom-up Broadband (BuB) branch of the \emph{Commons for Europe} (C4EU) European project aims to be a catalyzer that makes possible the collaboration among these groups and establishes the necessary documentation, methodology, and good practices to ease and replicate BuB network deployments that ultimately benefit European citizens.
The present paper covers the project definitions and goals, as well as the pilot proposals identified during the first 6 months of the project.
In our future works, we will provide a more detailed description of the executed pilots an results of the project.

The main characteristic of BuB is that the beneficiaries of the network (individuals or organizations) play an active role in the network planning, funding, deployment and maintenance.

The remainder of this article is organized as follows.
The next section briefly reviews previous work.
Sec.~\ref{sec:commons} introduces the concept of common resources in computer networks.
An overview of the specific technologies considered in the C4EU project is presented in Sec.~\ref{sec:technologies}.
Then, Sec.~\ref{sec:bub} provides a definition of BuB.
The pilot oriented methodology that we use in the C4EU project is described in Sec.~\ref{sec:methodology}.
The different BuB pilot proposals that have been considered in the C4EU project are summarized in Sec.~\ref{sec:proposals}.
Finally, Sec.~\ref{sec:conclusion} concludes the paper.

\section{Related Work}

BuB pilots have their precedents in wireless community networks \cite{flickenger2001wcn,jain2003wcn}.
BuB inherits the spirit of collaboration and the principles of inclusiveness.
It is also related to Municipal WiFi initiatives \cite{jassem2010mwc}, in the sense that BuB initiatives are sometimes backed by local authorities as an effective way to promote the availability and use of data networks.

BuB also represents a shift from the traditional top-down deployments.
Alternative models for network deployments have been considered in the past in \cite{barcelo2008wom} and \cite{seraghiti2010upw}.

A guide for BuB deployments and illustrative case studies is presented in \cite{aichele2006wnd}.
The growth of \emph{guifi.net} and the consequent increment of Internet penetration in the area in which \emph{guifi.net} has developed is studied in \cite{oliver2010wca}.
\emph{guifi.net} is probably the world's largest community network with over 16,000 active nodes as of 2012.

An important aspect of BuB network is that the beneficiaries keep a tight control over the network.
Retaining the ownership and control of the network may have implications in strengthening the freedom of speech and other human rights \cite{baig2009dcc}.
\section{Common Resources}
\label{sec:commons}
Common resources, or \emph{commons} for short, are shared by communities for efficiency reasons.
The \emph{Commons for Europe} (C4EU) project explores the effectiveness of commons in two particular areas or branches: software and networking.
To be more specific, the interest of the software branch is on simple applications (such as web applications or mobile applications) that can be used by different cities to improve the quality of life of their citizens and make a better use of the public resources.
The spirit of the project is to encourage the participation and implication of the citizens in making their city a better place to live.

A recurrent example is the application \emph{adopt a hydrant} in which citizens take responsibility for shoveling out a fire hydrant after it snows.
This eases the task of firemen in case there is a fire in the neighbourhood.
The \emph{adopt a hydrant} application is one of the most emblematic achievements of the \emph{Code for America} initiative \cite{codeforamerica}.

The other aspect of interest in the C4EU project is Bottom-up Broadband (BuB).
The aim of BuB is to deploy networks and networking services that can be used as a common resource.
A paradigmatic success story of the use of commons in networking is the allocation of the Industrial, Scientific and Medical (ISM) radio bands.
It is mentioned in \cite{abramson2009asw} that Michael Marcus proposed in the eighties the establishment of unlicensed bands as a common resource.
The availability of these bands has spurred the growth of the wireless industry making possible the IEEE 802.11 (WiFi) technologies that are in widespread use nowadays.

Note that this particular example is exposed to \emph{the tragedy of the commons} \cite{hardin1968tc,feeny1990tct}.
The tragedy of the commons is a well studied situation in which the members of a community over-exploit a common resource and substantially reduce its value.
In the case of ISM bands, if the number of transmissions grows too high, the resultant interference will render the network unusable.

Even though the tragedy of the commons may occur in certain busy locations at busy hours, it is generally considered that the advantages of having a commons spectrum band clearly outweigh the disadvantages.

Another good example of commons in networking is optical fiber deployment.
Deploying the cable can have a high cost, but every cable has multitude (e.g., 96) fibers.
Differently from WiFi technology, fibre optics do not place practical limitations in terms of available bandwidth.
After deploying the cable, some of these fibers may remain unused and the owner may decide to offer them as a common resource in order to facilitate the growth of the network.

A recurrent statement in networking is that the value of a network is proportional to the square of the number of nodes.
This somewhat controversial rule of thumb is known as Metcalfe's Law.
Despite the controversy, there is a general agreement that the value of the networks grows super-linearly with the number of nodes \cite{odlyzko2005rml}.
This means that, in many situations, adding new nodes to a network represents a win-win situation for all the participants.
If this is the case, the owners of the network may be willing to offer their own network resources as commons, as the value of their network will increase when others connect to it.

The BuB branch of the C4EU project is devoted to the exploration of networking initiatives in which different individuals and organizations team up to create and extend data networks using common resources.

The idea of commons can take many different forms: common infrastructure, common cabling, common hardware, common bandwidth, common spectrum, common knowledge, etc.

\section{Technologies of interest}
\label{sec:technologies}

In the project, four different technologies have been considered: WiFi, fibre optics, sensors and SuperWiFi.
WiFi technology has allowed the deployment of large networks at a low cost.
WiFi equipment is widely available and hackers have used it to interconnect nodes that carry the network traffic over long distances.
If the density of nodes is low (rural areas), highly directive links can be used to cover long distances.
When the density of nodes is higher, the nodes can use less directive antennas to reach and receive data from several nodes.

Radio communication is subject to interference and weather conditions.
Furthermore, since the ISM spectrum is a finite resource, there is a natural limit in the amount of bandwidth available with this technology.

For this reason, fibre optics is also a technology under consideration in BuB deployments.
The bandwidth available using optical links is orders of magnitude larger than that of WiFi.
Furthermore, links are more stable and are not affected by the interference or the weather.
Compared to wireless, deploying a fibre optics link is much more expensive.
Nevertheless, if this link can be used as a common resource, fibre optics can be economically viable.
Furthermore, it may be the case that there are already publicly owned fibers that can be used as a common resource.

In the BuB parlance, fiber deployments are usually termed FFTF (fiber from the farm \cite{roca2011fft}) or FFTH (fiber from the home).
These names emphasize the bottom-up nature of these deployment.

Another technology under consideration is sensor networks.
Quite often, sensor nodes are battery-powered devices with limited communication capabilities.
Strictly speaking, they do not represent a broadband technology.
Nevertheless, given the possibility and the interest by some organizations to deploy their own sensor networks in a bottom-up fashion, we have included them in our study of bottom-up networking.

The novelty of sensor technology makes it difficult to find existing success models in which it is used as commons.
The C4EU project should be helpful in defining these models.
As an example, an entity interested in gathering street parking information could collaborate with an entity interested in gathering pollution data and deploy the network as a common resource.
An interesting aspect regarding sensor networks is that the data itself can be offered as a common resource to allow independent developers to built applications using that data.

Finally, the last considered technology is SuperWiFi, which attempts to reproduce the success of WiFi at a different frequency band.
In particular, SuperWifi operates at lower frequencies (TV white spaces).
The idea is to use the spectrum unused by TV channels as commons.
The problem is that the equipment intended for this band must comply with very strict requirements to prevent the interference with licensed users.

The advantages of SuperWiFi compared to regular WiFi is that at lower frequencies we have larger coverage areas and better in-building penetration.
Increased coverage has also the downside of higher interference radius.

\begin{table}[!t]
\renewcommand{\arraystretch}{1.3}
\caption{Technologies under consideration}
\label{tab:technologies}
\centering
\begin{tabular}{|c|p{5cm}|}
\hline
Technology & Characteristics \\
\hline
Fibre Optics & Mature technology, wired, very high throughput, relatively expensive, does not create nor suffer interference, reliable. \\
WiFi & Mature technology, wireless, high throughput, more economic than fibre, limited by interference and spectrum saturation. \\
Sensors & New technology, wireless, low throughput (for battery-powered devices), open data. \\
Super-WiFi & Future technology, wireless, medium throughput, longer propagation distance and better penetration compared to WiFi, co-existence with incumbent networks.\\
\hline
\end{tabular}
\end{table}

\begin{figure}[!t]
\centering
\includegraphics[width=\linewidth]{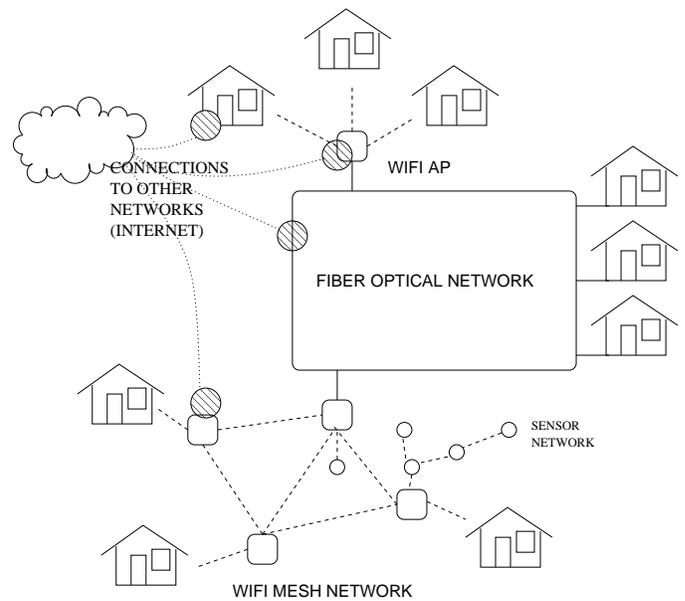}
\caption{An hybrid BuB deployment combining different technologies.}
\label{fig:hybrid}
\end{figure}

The main characteristics of the technologies considered in the C4EU project are summarized in Table~\ref{tab:technologies}.
Some of the considered deployments include more than one of the above mentioned technologies.
An example of a BuB network combining fiber optics, WiFi, mesh and sensor technologies is illustrated in Fig.~\ref{fig:hybrid}.
The sensors benefit from the bandwidth and coverage of WiFi and mesh technologies to send the sensed data and the wireless traffic is later aggregated in high-speed fibre optics links.

Besides technology, there is also an interest in the C4EU project to explore regulatory issues.
Regulation can deeply impact the ways in which BuB can be used and deployed.
Our intention is to obtain results that can help policymakers to decide which are the regulatory directions that lead to a greater benefit for the population.

\section{Bottom-up Broadband}
\label{sec:bub}

BuB networks are characterized by the implication of the beneficiaries in the planning, funding, deployment and maintenance of the network.
The beneficiaries do not necessarily have to be the end users.
The broad term beneficiaries refers to any person or organization that is interested in the availability of broadband services.

An illustrative example of this situation is a touristic destination that decides to offer public broadband services to visitors.
In this case the users are tourists, which are not the promoters of the bottom-up network.
Nevertheless, by being able to share their experiences online, these tourists will help in publicizing the touristic destination.
And this publicity benefits the promoters of the network.

Another example is one of the pilot proposals described in \cite{barcelo2012bpp} in which the promoter of the network is a football club.
The users of the network will be the fans and the club will benefit from the network as it will help to strengthen the community that ultimately supports the club.

The promoters of the network can be individuals or organizations.
And these organizations can be private companies, public institutions, non-governmental organizations, etc.  
It is also perfectly possible that the backers of BuB network represent a heterogeneous group, that share a common interest in the deployment of the network.

Deploying a broadband network is not an easy task.
It requires substantial amounts of money, time and know-how.
One of the goals of the C4EU project is to help to organize, publicize and transfer the  BuB know-how among those pilots that already have the money and time required to succeed.

Some people might be reluctant to accept that it is possible to deploy broadband networks following a collaborative approach.
For this reason, the first BuB initiatives should be oriented to achieve \emph{quick wins}.
A \emph{quick win} is a small, not particularly ambitious, project that can be successfully completed in a short time.
The completion of a \emph{quick win} will provide the necessary confidence for the involvement in larger, more complex projects.

\begin{figure}[!t]
\centering
\includegraphics[width=\linewidth]{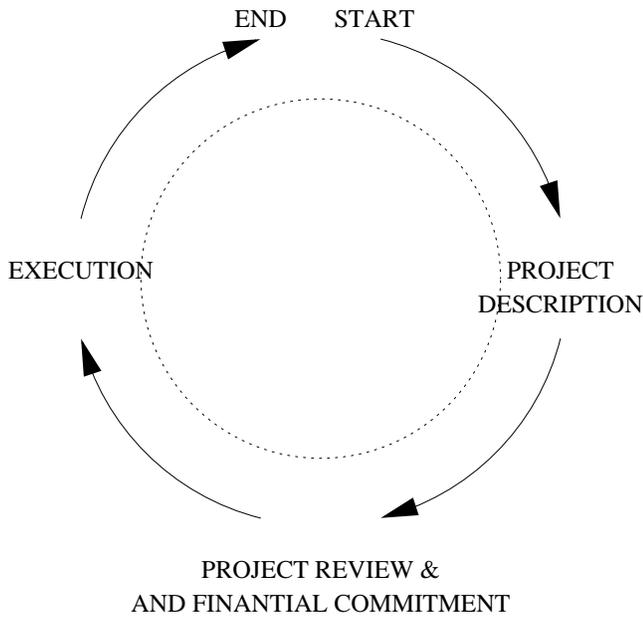}
\caption{Success by recursion.}
\label{fig:success_wheel}
\end{figure}

This approach has been termed \emph{success by recursion} and it is graphically represented in Fig.~\ref{fig:success_wheel}.
It differentiates three main steps for each project: description, review \& financial commitment, and execution.
The emphasis is placed in the fact that the successful completion of one project triggers the beginning of another one.

\section{Methodology}
\label{sec:methodology}

The planning of the C4EU project includes two differentiated phases: BuB pilots and BuB long term sustainability.
In the first phase selected BuB pilots will be executed with the help and under the supervision of the participants of the C4EU project.
The intention is to run several pilots using different technologies in different cities across Europe.
Each of these pilots should have a strong backing from local institutions and local champions that should make it possible for the pilots to progress autonomously after the first guided steps.

The second phase, which is much more ambitious, is the creation of a community that can continue the work started by the C4EU project in the long term.
This community, with the help of the know-how gathered during the C4EU pilot, should be able to continue to provide advice and support for the creation of new BuB pilots after the C4EU project has finished.

The first step of the project has been the dissemination of a call for pilot proposal in which the participants of the project have identified and documented potential pilots to be executed within the C4EU project.

\begin{figure}[!t]
\centering
\includegraphics[width=\linewidth]{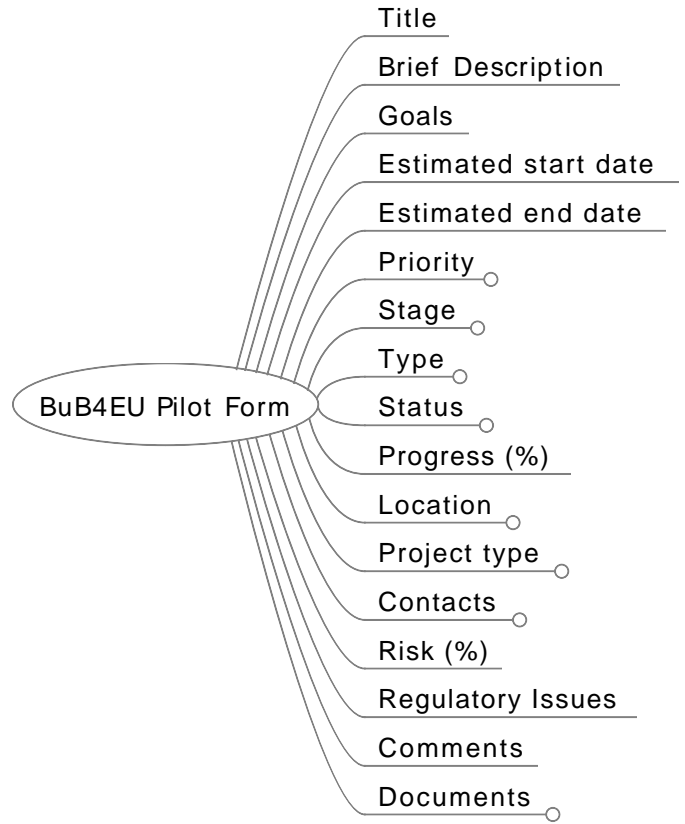}
\caption{Pilot proposal template.}
\label{fig:BuB4EU_pilot_template_upf}
\end{figure}

To systematize the information gathering process, we have created a pilot proposal template that contains the items represented in Fig.~\ref{fig:BuB4EU_pilot_template_upf}.  
More details about the exact meaning of each item are offered in \cite{barcelo2012bpp}.

A second call for pilots is programmed for the second year of the project. 
The call for pilots and the gathered information offers a picture of the state of Bottom-up Broadband initiatives in Europe.
More information about the C4EU project, the call for pilots and the pilot proposals can be found in \cite{barcelo2012bpp}.

\section{BuB Pilot Proposals}
\label{sec:proposals}

This section offers a very short summary of the twelve pilot proposals that we have received in answer to the call for pilots.
Note that in this paper we present only a very brief excerpt of every BuB proposal for conciseness. 
More details are available in the technical project \cite{barcelo2012bpp}.
Table~\ref{tab:pilot_opportunities} summarizes the different pilot proposals and classifies them by technology.

\begin{table}[!t]
\renewcommand{\arraystretch}{1.3}
\caption{Pilot opportunities}
\label{tab:pilot_opportunities}
\centering
\begin{tabular}{|c||c||c|}
\hline
Wifi/SuperWifi & Fiber & Sensor\\
\hline
FCMOSNET & FCMOSNET & \\
NQN & NQN & \\
Hulme High Street & Hulme High Street & \\
CAC & & \\
& PRBB & \\
OpenWisp& & \\
EuropeWIFI& & \\
& Gurb & \\
& Rubi & \\
Vic & Vic & \\
& & sensorWIFI\\
& & Open Sensor Network\\
\hline
\end{tabular}
\end{table}

\subsection{FCMOSNET}
FCMOSNET stands for FC United Moston Community Stadium Network and it has been proposed by the Manchester Digital Development Agenda (MDDA).
The goal is to create a BuB network that combines optical fibre and WiFi technologies around a new stadium built by the ``Football Club United of Manchester''.
This network should provide low-cost (or free) WiFi and low-cost fibre Internet connections to the members of the community.

\subsection{NQN}
NQN stands for Northern Quarter Net and has also been proposed by the MDDA.
The goal is to offer a public WiFi service which uses low-cost equipment and an existing high-speed fibre Internet connection.
NQN is being constructed as an Industrial Provident Society (IPS), which is an organization conducting an industry, business or trade in the form of a cooperative or for the benefit of a community.

\subsection{Hulme High Street}
This is the third pilot proposal by the MDDA and it is very similar in nature to the above mentioned NQN.
The goal is to deploy a community-led and owned broadband network that combines fibre optics and WiFi technology.

\subsection{CAC}
CAC stands for Catalonia's Audiovisual Council and the goal of this pilot is to use the available spectrum from the digital dividend to broadcast IPTV content using SuperWiFi.
The broadcast will have a limited geographical coverage and it is aimed to serve the interests of local communities.
This pilot has been proposed by Universitat Pompeu Fabra.

\subsection{PRBB}
PRBB is the Barcelona's Biomedical Research Parc, which hosts a combination of public laboratories and private spin-offs.
While the public laboratories are entitled to access a local public research network, the private spin-offs need an alternative to reach the Internet.
The availability of public fiber deployments in the neighborhood of the park makes it possible to implement a commons model to provide high-speed connectivity.
This pilot has been proposed by \emph{guifi.net}.

\subsection{OpenWisp}
OpenWisp is a project initiated by the Province of Rome to extend the opportunities of the citizens to access broadband Internet.
This service is publicized with the name \emph{ProvinciaWiFi} and makes use of free software tools and open standards combined with low-cost hardware.
It has deployed several hundreds of access points that offer access to registered users in public locations.
The platform is offered as a common resource for others to replicate and collaborate.
This project has been jointly proposed by Provincia di Roma and CASPUR.

\subsection{EuropeWiFi}
The idea of this project is to replicate the success model of ProvinciaWiFi and construct a RADIUS hierarchy similar to the existing EduRoam to make it possible for the European citizens to use the public Internet access when they visit other cities or countries.
The goal is to build a pan-European public WiFi solution relying on free software, open standards and low-cost hardware.
This project is also promoted by CASPUR and Provincia di Roma.

\subsection{Gurb}
Gurb is a village that has already carried out a first phase of BuB fibre optics deployment.
This pilot represents a second phase of the deployment that re-uses municipality-owned infrastructure.
These initiatives are described with the acronym FFTF which means Fiber From The Farm.
The emphasis is placed in the fact that the deployment starts from the farm, making it clear that it is a truly BuB initiative.
The organization that orchestrates the deployment is called GurbTec.

\subsection{Rubi}
Rubi is a medium-sized city (ca. 75,000 inhabitants) and there is interest in starting a BuB deployment to cover both industrial and residential areas.
This will be a Fiber From The Home (FFTH) deployment and the intention is to start with a small project that can be executed in a few months to provide a \emph{quick win} and gather support for new phases of the project that further extend the network.
The promoters include an heterogeneous group of citizens, business and public administration.

\subsection{Vic}
Vic is a small city (ca. 40,000 inhabitants) interested in a BuB FFTH deployment to offer services to citizens, schools and businesses.
The promoters are two groups called Gaufix and Gurbtec.

\subsection{SensorWifi}
The idea of this pilot is to re-use the existing coverage offered by ProvinciaWiFi to gather data from sensors deployed across the city.
In this case, the WiFi infrastructure is the commons resource to be shared by different services and applications.
This pilot proposal has been suggested by CASPUR and Provincia di Roma.

\subsection{Open Sensor Network}
The goal of this pilot is to construct a platform to collect and share real-time information gathered by sensors.
If this information is publicly available, it can be re-used by application developers to create imaginative applications (typically web applications or mobile applications) that will be useful to the citizens.

\section{Pilot Planning Template}
\label{sec:planning}
In this section we include and example pilot planning template for a BuB deployment. 
Obviously, every BuB initiative is different and the example template will need to be adapted to the requirements of each deployment.
Nevertheless, this template can be used as a guideline and checklist to make sure that no important step has been skipped.
A more detailed plan which includes the duration of each task and inter-dependencies in provided in the technical report \cite{barcelo2012bpp}.

\begin{itemize}
  \item{Warm-up:} 
  The first task involves gathering information regarding the deployment, a first draft of a high-level deployment plan and commitment from the leaders of the participating organizations.
  This previous initial agreement is a fundamental step, and a requisite to start working on the details of the deployment.
  \item{Project plan:}
  The second task is to prepare a detailed working plan.
  \begin{itemize}
    \item{Scope definition:}
    At this point, specific goals for the project will be settled.
    In general, it is advisable to aim for a \emph{quick win} by setting realistic goals that can be achieved in a short period of time.
    \item{Project development:}
    All the necessary steps to reach the defined goals should be carefully documented.
    This documentation will be used as an input for the next task.
  \end{itemize}
  \item{Project review \& commitment:}
  At this point, the project plan will be either accepted for execution or rejected.
  An acceptance means continuing with the next task while a rejection implies moving back to the Project Plan task to prepare a refined version of the documentation that can be approved in a subsequent round.
  \begin{itemize}
    \item{Identify stakeholders:}
    At this stage, all the people involved in the deployment should be identified.
    Participants include the users, the professionals that work on the deployment and the service providers and the investors.
    The acceptance of the project is confirmed by the financial commitment from part of the investors.
    After the commitment, the project is ready for execution.
  \end{itemize}
  \item{Execution:}
  In this task is where the actual deployment occurs.
  \begin{itemize}
    \item{Provisioning:}
    Obtaining all the required equipment and tools for the deployment.
    It is recommended to test and configure the equipment at this stage in a controlled environment.
    \item{Deployment:}
    Installation of cables, antennas and networking devices.
    \item{Wholesale access:}
    The working network is connected to an upstream provider to the Internet.
    \item{Legal \& compliance:}
    Some deployments may require notification to an authority or regulator, or other legal and bureaucratic procedures.
  \end{itemize}
  \item{Start operations:}
  At this point the network is ready to start operations.
  \item{Project Management:}
  This task spans throughout all the project and includes the supervision and necessary corrections to steer the direction of the project.
\end{itemize}

\section{Conclusion}
\label{sec:conclusion}

In this paper we have offered an overview of the BuB branch of the C4EU European project.
First, we have introduced the concept of commons, which are common resources shared by a community.
Our main interest is the use of commons in the context of data networks.
To be more specific, four technologies are considered in the C4EU project: WiFi, fibre optics, sensor networks and Super-WiFi.

Then, we have provided a definition of BuB.
The key concept is that the beneficiaries of the network get involved in all the aspects of the network: design, planning, funding, deployment and maintenance.

Regarding the methodology, the interest is on pilot proposal and execution for the first years of the project, and on long-term sustainability aspects at the end of the project.
We have summarized the pilot proposals that we have received as an answer to a first call for pilots and described a basic outline for pilot planning and execution.

To sum up, the BuB branch of the C4EU project is hands-on approach to understand, define and document bottom-up data networks initiatives that rely heavily on common shared resources.


\section*{Acknowledgment}

This work has been partially funded by the European Commission (grant CIP-ICT PSP-2011-5).
The views expressed in this technical report are solely those of the authors and do not represent the views of the European Commission.



\bibliographystyle{IEEEtran}
\bibliography{IEEEabrv,my_bib}
%

%

\end{document}